**"Response of RAW 264.7 and J774A.1 macrophages to particles and nanoparticles of a mesoporous bioactive glass: A comparative study"**

M.J. Feito[a,*], L. Casarrubios[a], M. Oñaderra[a], M. Gómez-Duro[a], P. Arribas[a], A. Polo-Montalvo[a], M. Vallet-Regí[b,c,*], D. Arcos[b,c], M.T. Portolés[a,c,*]

[a] Departamento de Bioquímica y Biología Molecular, Facultad de Ciencias Químicas, Universidad Complutense de Madrid, Instituto de Investigación Sanitaria del Hospital Clínico San Carlos (IdISSC), 28040 Madrid, Spain.

[b] Departamento de Química en Ciencias Farmacéuticas, Facultad de Farmacia, Universidad Complutense de Madrid, Instituto de Investigación Sanitaria Hospital 12 de Octubre i+12, Plaza Ramón y Cajal s/n, 28040 Madrid, Spain.

[c] CIBER de Bioingeniería, Biomateriales y Nanomedicina, CIBER-BBN, 28040 Madrid, Spain.

* Corresponding authors: mjfeito@ucm.es, vallet@ucm.es, portoles@quim.ucm.es

Total number of words 7,946
1 Scheme and 7 Figures

**ABSTRACT**

Mesoporous bioactive glasses (MBGs) are bioceramics designed to induce bone tissue regeneration and very useful materials with the ability to act as drug delivery systems. MBGs can be implanted in contact with bone tissue in different ways, as particulate material, in 3D scaffolds or as nanospheres. In this work, we assessed the effects of particles of mesoporous bioactive glass MBG-75S and mesoporous nanospheres NanoMBG-75S on RAW 264.7 and J774A.1 macrophages, which present different sensitivity and are considered as ideal models for the study of innate immune response. After evaluating several cellular parameters (morphology, size, complexity, proliferation, cell cycle and intracellular content of reactive oxygen species), the action of MBG-75S particles and NanoMBG-75S on the polarization of these macrophages towards the pro-inflammatory (M1) or reparative (M2) phenotype was determined by the expression of specific M1 (CD80) and M2 (CD206, CD163) markers. We previously measured the adsorption of albumin and fibrinogen on MBG-75S particles and the production of pro-inflammatory cytokines as TNF-α and IL-6 by macrophages in response to these particles. This comparative study demonstrates that particles of mesoporous bioactive glass MBG-75S and mesoporous nanospheres NanoMBG-75S allow the appropriated development and function of RAW 264.7 and J774A.1 macrophages and do not induce polarization towards the M1 pro-inflammatory phenotype. Therefore, considering that these mesoporous biomaterials offer the possibility of loading drugs into their pores, the results obtained indicate their high potential for use as drug-delivery systems in bone repair and osteoporosis treatments without triggering an adverse inflammatory response.

# 1. INTRODUCTION

The increase in life expectancy in modern societies highlights a higher prevalence of age-related bone diseases such as osteoporosis, bone cancer and bone infections [1]. Regarding biomaterials used for the repair of bone defects, these are implantable materials designed to contact with living tissues, trigger bone formation and produce an adequate immune response. In this context, the implantation of any biomaterial in the organism triggers different events, particularly, in the first stages after biomaterial-tissue contact, proteins from physiological fluids and blood adsorb to the biomaterial surface and activate the coagulation cascade, complement system, platelet adhesion as well as cells involved in the immune response [2,3]. Coagulation together with platelet adhesion and activation, constitute the main events in surface-induced thrombosis, which is the most critical limitation of biomaterials that may be in contact with blood [4]. Thrombosis is a multifactorial pathology, resulting from the interaction of multiple factors leading to an imbalance between the prothrombotic, anticoagulant and thrombolytic processes. Hypoalbuminemia is another factor related to hyperaggregability [5]. Fibrinogen, albumin, immunoglobulins, vitronectin and apolipoproteins are usually present in significant amounts on the adsorption surfaces of different biomaterials [6]. Fibrinogen is involved in platelet adhesion to biomaterials and therefore, the biomaterial surfaces that avoid fibrinogen adsorption should be more hemocompatible. Fibrinogen binds to an integrin-type receptor, GP IIb/IIIa which bridges platelets, allowing the formation of platelet plugs to stop bleeding in damaged arteries and reducing the autoactivation of the intrinsic coagulation pathway [6-8]. Plasma albumin is a suitable parameter to assess the risk of thrombosis since concentrations below 2 mg/dL are associated with a high risk of thrombotic complications [9]. Albumin constitutes approximately 50% by weight of the plasma protein content but is generally low represented in the plasma protein layer adsorbed on the surface of biomaterials [10]. Sivaraman and Latour showed that platelets adhere to albumin layers when this protein is unfolded exposing amino acid sequences that bind to integrin-type receptors, GPIIb/IIIa, which are the same ones to which the adsorbed fibrinogen binds [11].

Regarding the activation of the immune system cells by biomaterials, the evaluation of these effects are essential aspects of biocompatibility assessment, as a prior process to implantation in an organism [12-14]. Thus, the immune system is activated by the implantation of any foreign body, triggering a cascade of inflammatory processes such as activation of macrophages, formation of fibrous capsules and cytokine release around the implant [15]. Neutrophils, monocytes, and macrophages are essential cells of the innate immune response, involved in phagocytosis and producing reactive oxygen species, antimicrobial peptides, and inflammatory mediators [14]. Recently, it has been shown that macrophages would be beneficial for biomaterial integration after implantation due to the remarkable macrophage plasticity [16]. A broad phenotypic spectrum of macrophages has been described between two extremes identified as pro-inflammatory M1 and reparative M2 macrophages, characterized by the expression of specific cell surface markers and the secretion of different cytokines [17]. Whereas M1 macrophages release pro-inflammatory cytokines such as IL-1β, IL-6, IL-12 and TNF-α, M2 macrophages produce anti-inflammatory factors such as IL10, IL-1 receptor type α and TGF-β, inducing activation of the Th2 immune response, which is considered a key process for tissue regeneration [18-20]. Interestingly, M2 macrophages are often described as the angiogenic phenotype since they also produce potentially angiogenic factors such as VEGF, FGF-β, and TGF-β [21,22].

Mesoporous bioactive glasses (MBGs) are bioceramics designed to induce bone tissue regeneration and very useful materials with the ability to act as drug delivery systems due to their highly ordered mesoporous structure [23-26]. The large surface area of MBGs, their pore volume and open channel morphology, facilitate dissolution and ion exchange with the surrounding fluids, producing an increase in calcium and phosphate ion concentrations in the areas between bone and implant, facilitating the formation of new apatite layers [27] and, therefore, exhibiting high bioactivity [28]. In addition, these materials offer the possibility of loading into their pores anti-osteoporotic drugs, antibiotics, and other agents for subsequent release [25,28-32]. With these objectives, MBGs can be implanted in contact with bone tissue in different ways, as particulate material, in 3D scaffolds or as nanospheres. In this sense, mesoporous nanospheres are recently being prepared as very promising nanomaterials due to their great potential as intracellular drug delivery systems, together with their low toxicity and high biocompatibility [33-36]. Recent studies have demonstrated the usefulness of mesoporous nanospheres loaded with ipriflavone for the intracellular release of this drug in pre-osteoblasts in monoculture and in osteoblasts/osteoclasts coculture, evidencing their potential for osteoporosis and periodontal disease treatment [33,34].

In this work, we assessed the effects of particles of mesoporous bioactive glass MBG-75S and mesoporous nanospheres NanoMBG-75S on RAW 264.7 and J774A.1 macrophages, which present different sensitivity and are considered as ideal models for the study of innate immune response. After evaluating several cellular parameters (morphology, size, complexity, proliferation, cell cycle and intracellular content of reactive oxygen species), the action of MBG-75S particles and NanoMBG-75S on the polarization of these macrophages towards the pro-inflammatory (M1) or reparative (M2) phenotype was determined by the expression of specific M1 (CD80) and M2 (CD206,

CD163) markers. We previously measured the adsorption of albumin and fibrinogen on MBG-75S particles and the production of pro-inflammatory cytokines as TNF-α and IL-6 by macrophages in response to these particles. Regarding the studies with MBG-75S particles and considering the important role of the ions released by this particulate material [25], two different experimental strategies were followed to expose macrophages directly or indirectly (using semipermeable transwells) to MBG-75S particles in order to differentiate the action of these ionic products from the effects produced by the direct contact of the cells with the biomaterial. On the other hand, NanoMBG-75S were added directly to the culture to be incorporated intracellularly by macrophages. These three experimental conditions, shown in Scheme 1, were used with RAW 264.7 and J774A.1 macrophages in parallel.

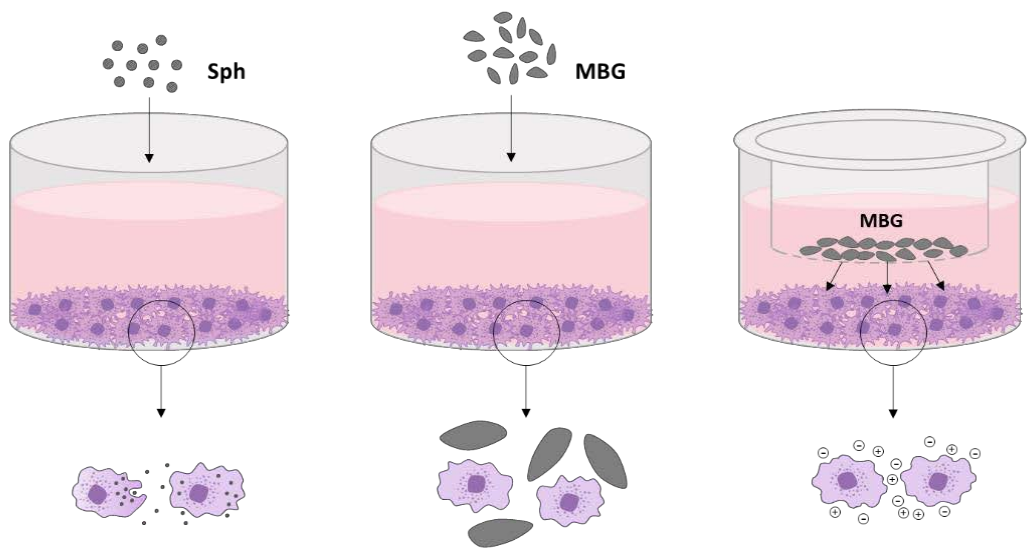

**Scheme 1.** Direct and indirect treatment of cultured macrophages (RAW 264.7 or J774A.1 cells) with NanoMBG-75S nanoparticles (nanospheres, Sph) or MBG-75S particles (MBG).

The final objective of this study is to compare the effect of MBG-75S particles and NanoMBG-75S, potentially useful for bone regeneration and osteoporosis treatment, on two types of macrophage cell lines with different sensitivity, shedding light on the immune response to these biomaterials which must be known prior to their use *in vivo*.

## 2. MATERIALS AND METHODS

*2.1. Synthesis of materials.*

*2.1.a. Synthesis of mesoporous bioactive glass particles (MBG-75S)*

A mesoporous bioactive glass with nominal composition $75SiO_2$-$20CaO$-$5P_2O_5$ (% mol) was synthesized using the evaporation induced self-assembly method. For this aim, a non-ionic triblock copolymer $(PEO)100$-$(PPO)65$-$(PEO)100$ [PEO is poly(ethylene oxide) and PPO is poly(propylene oxide)], pluronic F127, as structure directing agent. Briefly, 4 g of Pluronic F127 were dissolved in a solution of 2 mL of HCl 1M and 60 g of ethanol. Afterwards, 7.70 mL of tetraethylortosilane, $Si(OC_2H_5)_4$, TEOS, 1.05 mL of triethylphosphate, $P(OC_2H_5)_3$, TEP and 2.93 mg of calcium nitrate tetrahydrate, $Ca(NO_3)_2 \cdot 4H_2O$, were added at three hours intervals. The mixture was magnetically stirred for 24 h and then poured into Petri dishes to evaporate at 30 °C for 7 days. After this period, the material was removed from the dishes as transparent and homogeneous membranes, which were heated at 700 ºC for 3 h in order to remove nitrates and organic compounds. Finally, the resulting solid was gently milled and sieved, collecting the particles ranging in size between 2 and 50 micrometers.

*2.1.b. Synthesis of mesoporous bioactive nanospheres (NanoMBG-75S)*

NanoMBGs were prepared using a dual soft template strategy to obtain SiO2-CaO nanospheres comprising a large hollow core and radially distributed channels in the shell [33]. For this aim, 80 mg of poly(styrene)-block-poly(acrylic acid) (PS-b-PAA) with average Mw = 38,000 Da were dissolved in 18 mL of tetrahydrofuran (THF) at room temperature under magnetic stirring (solution 1). Furthermore, 160 mg of hexadecyltrimethylammonium bromide (CTAB) were dissolved in 7.4 mL of D.I. water and 2.4 mL of ammonia (28% w/v) and stirred at 37 °C at 100 rpm in an incubator to avoid foaming of the solution (solution 2). After complete dissolution of both reactants, solution 1 was poured into solution 2 under vigorous stirring for 20 min. Thereafter the inorganic sources were added dropwise and step by step (25 μL of triethyl phosphate, TEP, in 1.6 mL ethanol, 125 mg of $Ca(NO_3)_2 \cdot 4H_2O$ in 1.6 mL of water and 0.52 of tetraethyl orthosilicate, TEOS, in 1.6 mL ethanol) at 20 min intervals. The mixture was covered with parafilm to avoid THF evaporation and stirred for 24 h at room temperature. Finally, the product was collected by centrifugation at 10,000× g rpm (g = 16.466) for 10 min and washed several times with an ethanol:water (1:1) solution. The resulting powder was dried and subsequently calcined at 550 °C for 4 h to remove the organic templates.

*2.2. Characterization of materials.*

MBG-75S particles and NanoMBG-75S nanospheres were studied by transmission electron microscopy and chemically analyzed be energy dispersed X-ray spectroscopy (TEM/EDX) using a using a JEOL-1400 microscope operating at 300 kV (Cs 0.6 mm, resolution 1.7 Å). Chemical composition was determined by EDX measurements during TEM experiments.

Textural properties were studied by means of nitrogen adsorption analysis using an 3Flex (Micromeritics) equipment. For this aim, MBG-75S particles and NanoMBG-75S nanospheres were degassed at 150 °C for 15 h. Fourier-transform infrared spectroscopy (FT-IR) was carried out using a Nicolet Magma IR 550 spectrometer

*2.3. Adsorption of bovine serum albumin (BSA) on particles of MBG-75S and Si substituted hydroxyapatite (SiHA)*

Adsorption of bovine serum albumin (BSA) was carried out with 10 mg of particles of MBG-75S and two Si substituted hydroxyapatites, previously treated at either 700 °C (nanocrystalline) or 1250 °C (crystalline), in order to compare this process on biomaterials with different microstructural properties. These samples were incubated with a solution of BSA (Sigma Chemical Company, St. Louis, MO, USA) at 1.35 mg/mL in PBS and kept under gentle shaking for 4 h at 37 °C. No significant difference was detected between the amount of adsorbed protein after 4 h and 24 h of incubation at 37 ºC. Controls with the same BSA concentration but without material were performed in parallel. After 4 hours, the quantity of adsorbed protein was measured by UV-vis spectroscopy and calculated as the difference in protein concentration before and after BSA adsorption remaining in the supernatant. The extinction coefficient used for these experiments was of 0.667 (279 nm, 1 cm and 0.1% solution of BSA).

*2.4. Adsorption of bovine fibrinogen on particles of MBG-75S and Si substituted hydroxyapatite (SiHA)*

Adsorption of fibrinogen (Sigma Chemical Company, St. Louis, MO, USA) was performed by incubation of this protein (1 mg/mL in PBS) at 37 °C for 4 h with 10 mg of particles of MBG-75S and two Si substituted hydroxyapatites, previously treated at either 700 °C (nanocrystalline) or 1250 °C (crystalline), in order to compare this process on biomaterials with different microstructural properties. The protein amount in the supernatants was measured by UV-vis spectroscopy, and the percentage of adsorbed fibrinogen was then calculated as the difference between the initial fibrinogen concentration and the fibrinogen remaining in the resulting supernatant.

*2.5. Culture of RAW 264.7 and J774A.1 macrophages*

Two murine macrophage cell lines were used for this study: RAW 264.7 and J774A.1 cells. A total number of $10 \times 10^4$ cells/mL were seeded on 6-well culture plates in 2 mL of Dulbecco's Modified Eagle's Medium (DMEM, Sigma Chemical Company, St. Louis, MO, USA) supplemented with 10% vol/vol fetal bovine serum (FBS, Gibco, BRL, UK), 200 µg/mL penicillin (BioWhittaker Europe, Verviers, Belgium), 200 µg/mL streptomycin (BioWhittaker Europe, Verviers, Belgium) and 1 mM L-glutamine (BioWhittaker Europe, Verviers, Belgium). The culture medium was changed after 24 hours to allow a correct cell adhesion and to eliminate the non-adherent cells. Then, 1 mg/mL of MBG-75S or 50 µg/mL of NanoMBG-75S were added and incubated for 48 hours. In parallel, to evaluate the effect of the ions released from MBG-75S on these macrophages, 1 mg/mL of this material was deposited on 6-well transwell inserts (0.4 µm pore size, Corning Inc., Corning, NY, USA), and covered with culture medium, to be in indirect contact with the cultured cells. Cultures were maintained in standard conditions, at 37ºC under a 5% $CO_2$ atmosphere. The attached macrophages were washed with PBS and harvested using cell scrapers. The proliferation of RAW 264.7 and J774A.1 macrophages was determined with a Neubauer chamber and compared to the initial concentration of seeded cells. Cell suspensions were centrifuged at 310 x g for 10 min and resuspended in a fresh medium for the analysis of different parameters (Sections 2.6–2.8). Control conditions in the absence of materials were performed in parallel. To ensure statistical significance, $10^4$ cells per sample were analyzed.

*2.6. Cell size and complexity*

After detachment of RAW 264.7 and J774A.1 macrophages, cell size and complexity were analyzed through the forward angle (FSC) and side angle (SSC) scatters, using a FACScalibur Becton Dickinson Flow Cytometer.

*2.7. Intracellular reactive oxygen species (ROS) content*

Cells were incubated with 100 µM of 2',7'-dichlorofuorescein diacetate (DCFH/DA, Serva, Heidelberg, Germany) for 30 min at 37 ºC. DCFH/DA can pass through the cell membrane and be deacetylated by cellular esterases to DCFH, which is quickly oxidized by the intracellular reactive oxygen species (ROS) present inside the cells to fluorescent DCF. To quantify intracellular ROS content, DCF fluorescence was excited at 488 nm with a 15 mW laser tuning of a FACScalibur Becton Dickinson Flow Cytometer and measured with a 530/30 filter.

*2.8. Morphological studies by confocal microscopy*

The effect of 1 mg/mL of MBG-75S in direct and indirect contact through transwell on macrophage morphology was observed after 48 hours of treatment by confocal microscopy, fixing the cells 24 h before microscopic observation. Confocal microscopy was carried out after using FITC phalloidin-rhodamine to stain actin filaments and DAPI to stain nucleus. For these studies, cells were seeded on glass coverslips and incubated in 6-well culture plates for 48 hours. After fixation with 4% paraformaldehyde in PBS for 10 min, samples were washed with PBS and permeabilized with 500 μL of Triton X-100 for 5 min. Samples were preincubated with PBS-BSA 1% for 20 min in order to avoid unspecified unions. Then cells were washed with PBS and incubated with 100 μL of phalloidin-rhodamine 1:40 for 20 min, washed with PBS and incubated with 100 μL of DAPI 3 μM for 5 min. Cells were examined using a Leica SP2 Confocal Laser Scanning Microscope, where the fluorescence of the dyes were excited at 540 nm (rhodamine) and 405 nm (DAPI) and measured at 565 nm (rhodamine) and 420/586 nm (DAPI).

*2.9. Characterization of M1 and M2 macrophage phenotypes*

The effects of 1 mg/mL of MBG-75S (in direct and indirect contact through transwell) and 50 μg/mL of NanoMBG-75S on RAW 264.7 and J744-A1 macrophage polarization towards the pro-inflammatory (M1) or reparative (M2) phenotype were evaluated by immunostaining and flow cytometry. After detachment and centrifugation at 310 x g, cells were washed and incubated in 45 μL of staining buffer (SB) (2% FBS in PBS, Sigma-Aldrich Corporation, St. Louis, MO, USA) with 10% of normal mouse serum inactivated (iNMS) for 10 min at 4 °C in order to block the Fc receptors of the plasma membrane to avoid unspecific unions.

In order to analyze the expression of CD80 as specific marker of cells polarized to M1 phenotype [36], samples were incubated with anti-mouse CD80 antibody conjugated with phycoerythrin (PE) (2.5 μg/mL, BioLegend, San Diego, California) in 1 mL of staining buffer (SB) (2% FBS in PBS, Sigma-Aldrich Corporation, St. Louis, MO, USA) for 30 min at 4° C in darkness

The percentage of cells polarized to M2 phenotype was measured by incubating the cell suspensions with anti-mouse CD206 antibody conjugated with fluorescein isothiocyanate (FITC, 2.5 μg/mL, BioLegend, San Diego, California) and an anti-mouse

CD163 antibody conjugated with Alexa Fluor 488 (A488) (2.5 µg/mL, BioLegend, San Diego, California) [37,38] for 30 min at 4 °C in darkness.

Cell suspensions were then washed twice with 500 µL of SB and resuspended in 300 µL of SB to analyze the samples in a FACSCalibur flow cytometer. Fluorescence of the fluorochromes was excited at 488 nm and measured at 585/42 nm for PE and 490/525 for FITC and A488. Each experiment was carried out three times and single representative experiments are displayed. At least $10^4$ cells were analyzed in each sample for statistical significance. Control conditions in the absence of materials were carried out in parallel.

*2.10. Detection of IL-6 and TNF-α as pro-inflammatory cytokines secreted by J774A.1 and RAW 264.7 macrophages*

The amounts of IL-6 and TNF-α secreted in the culture medium by RAW 264.7 and J774A.1 macrophages were quantified by enzyme-linked immunosorbent assay (ELISA, Gen-Probe, Diaclone) according to the manufacturer's instructions.

ELISA test consists of a sandwich in which plates precoated with a highly specific antibody for IL-6 or TNF-α, which entraps the soluble IL-6 or TNF-α molecules present in the culture medium of each sample after incubation. Later, a biotinylated secondary antibody was added, and the established unions were exposed in a colorimetric reaction with the addition of streptavidin-avidin conjugated with horseradish peroxidase. The quantity of the two cytokines were measured at 450 nm in an ELISA plate reader, with a sensitivity of 2 pg/mL and an inter-assay variation coefficient 3.6%. Control conditions were carried out in parallel and recombinant cytokine was used as standard.

*2.11. Statistical analysis*

Data are presented as mean ± standard deviation of three replicate experiments. Statistical evaluations were performed using the software Statistical Package for the Social Sciences 22$^{th}$ (SPSS Inc., Chicago, IL, USA). After having checked the normal distribution and homogeneity of variances, one-way analysis of variance (ANOVA) was performed to compare differences between groups. Then, the Scheffé and Games-Howell tests were used for post-hoc analysis to detect significant differences between the study groups. A level of p<0.05 was considered statistically significant: *p<0.05, ** p<0.005 and *** p<0.001.

## 3. RESULTS AND DISCUSSION

### *3.1.* **Characterization of mesoporous bioactive glass MBG-75S particles and NanoMBG-75S nanospheres**

MBG-75S particles and NanoMBG-75S nanospheres were studied by FT-IR spectroscopy to determine the different chemical groups in both materials (Figure 1.a). Both spectra show the characteristic absorption bands for a silica-based glass including the Si-O stretching band at 1050 cm$^{-1}$ and Si-O-Si bending mode a 550 cm$^{-1}$. MBG-75S particles also show a weak absorption singlet band at 610 cm$^{-1}$ that correspond to P-O bonds when phosphate groups are in an amorphous environment, as correspond to $P_2O_5$ in a glass material. This small band cannot be observed in the case of NanoMBG-75S nanospheres, pointing out that $P_2O_5$ could not be incorporated into the nanoparticles with the synthesis method used in this work.

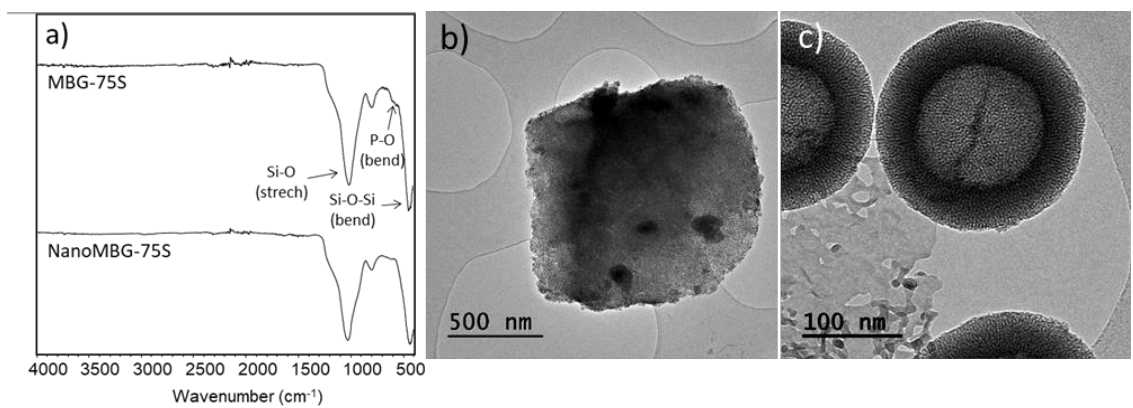

**Figure 1**. FT-IR spectra for MBG-75S particles and NanoMBG-75S nanospheres (a). TEM images obtained for MBG-75S (b) and NanoMBG-75S showing the double core shell porous structure (c).

TEM images evidence that MBG-75S particles have a grain size of 2 micrometers or larger, while showing an ordered mesoporous structure (Figure 1b). On the other hand, NanoMBG-75S nanospheres (Figure 1c) have a particle size of about 200 nanometers and are formed by an inner core with pores of about 100 nm, surrounded by a shell that exhibits a radial ordering of mesopores.

Figures 2a and 2b show the nitrogen adsorption isotherms for MBG-75S and NanoMBG-75S, respectively. Both materials show isotherms corresponding to mesoporous materials. However, the hysteresis loop for MBG-75S correspond to a H1 type loop pointing out the cylindrical and regular morphology of the mesopores of these particles.

On the other hand, the hysteresis loop for NanoMBG-75 exhibits an irregular shape pointing out that presence of at least two different porous systems in these nanoparticles. Figures 2c and 2d show the pore size distribution for both samples. MBG-75S particles show a single modal distribution centred at 6 nm. However, NanoMBG-75S shows a bimodal pore size distribution. The smaller one is centered at 2.3 nanometers and would correspond to the radial porosity of the nanospheres shell. The larger distribution is in the range of 30 micrometers or more and would correspond to the hollow core of the nanospheres, in agreement with the TEM observations. Table S2 (Supporting information) shows the textural parameters obtained by nitrogen adsorption analysis, indicating that both materials exhibit very high surface area values and porosities, as correspond to mesoporous materials.

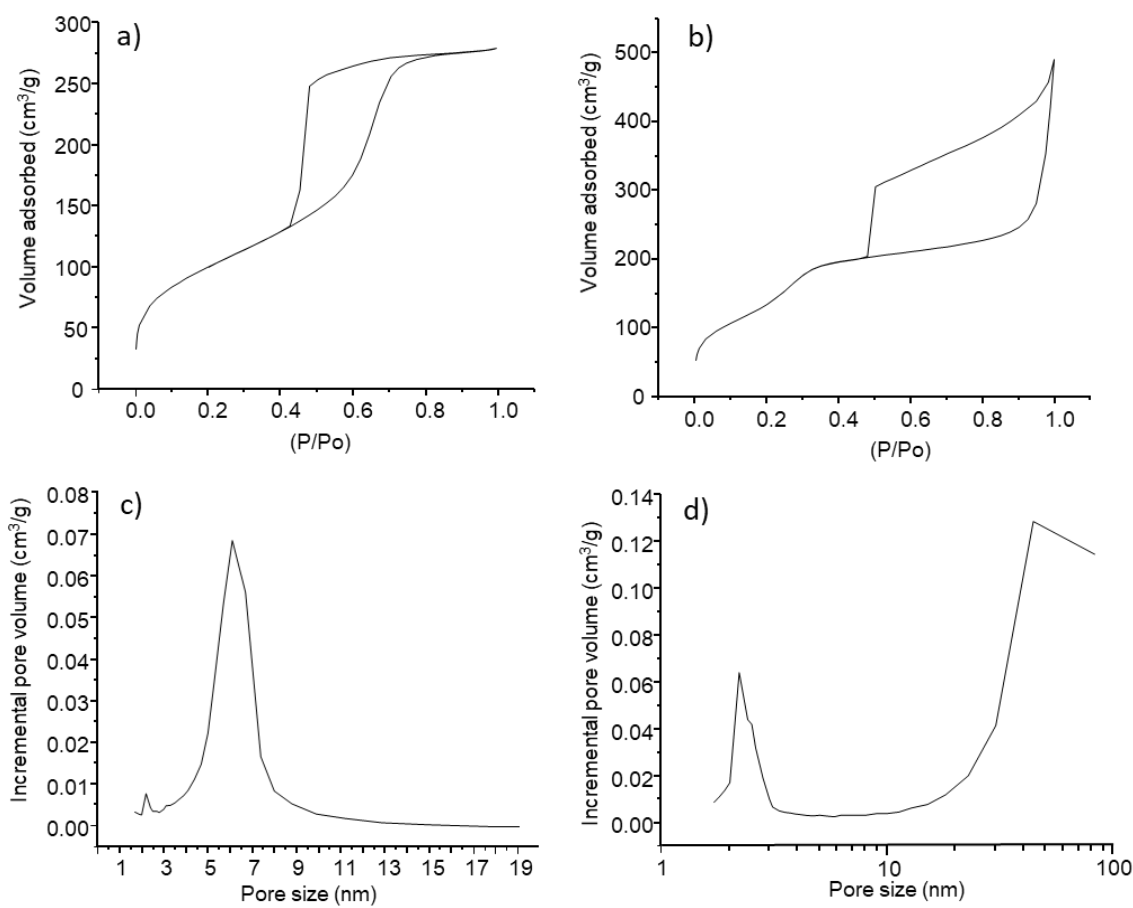

**Figure 2.** Nitrogen adsorption isotherms for MBG-75S particles (a) and NanoMBG75-S nanospheres (b); pore size distribution for MBG-75S particles (c) and NanoMBG75-S nanospheres (d). x axis of figure 2.d is represented in logarithmic ($\log_{10}$) scale for clarity purposes.

*3.2.* **Adsorption of serum albumin (BSA) and fibrinogen on particles of MBG-75S and Si substituted hydroxyapatite (SiHA)**

Biomaterial surfaces are rapidly covered *in vivo* with different host proteins [37,38] which can activate the coagulation cascade, complement system, platelet adhesion and immune cells [2,3]. In this context, it has been observed *in vitro* that denatured surface protein coated biomaterials lead to increased monocyte adhesion [39], which is the first step in a sequence of events that may ultimately culminate in adverse immune reactions *in vivo* [40]. In the present study, the adsorption of specific serum proteins as bovine serum albumin (BSA) and fibrinogen, was measured on particles of MBG-75S and on two types of Si substituted hydroxyapatites (SiHA), previously treated at 700 °C (nanocrystalline) and 1250 °C (crystalline), to compare this process on these three biomaterials with different microstructural properties.

The amounts of BSA and fibrinogen adsorbed on nanocrystalline SiHA were significantly higher than on crystalline SiHA ($p < 0.005$) (Figure 1S in Supporting information), in agreement with previous results obtained by our group with 3D scaffolds prepared with these two hydroxyapatites [41]. If we compare the results obtained with SiHA and MBG-75S materials, we can observe that the adsorption of fibrinogen on MBG-75S particles presents an intermediate value between the values obtained with these two hydroxyapatites, but nevertheless, the adsorption of BSA on MBG-75S material was significantly higher than on the two SiHA. This BSA adsorption result obtained with MBG-75S, can be considered beneficial due to the potential of albumin to reduce many biological interactions with surfaces, minimizing opsonization or inflammation and decreasing thrombogenicity, platelet adhesion and bacterial adhesion, thus preventing potential infection [38, 39]. Thus, albumin adsorption on this material might be associated with improved hemocompatibility. In contrast to albumin, fibrinogen adsorption is normally related to platelet adhesion and activation, leading to thrombus formation [4,5,42].

*3.3.* **Effects of direct contact of RAW 264.7 and J774A.1 macrophages with MBG-75S particles on TNFα and IL-6 secretion**

The evaluation the immune system activation by the biomaterials to be implanted is an essential aspect for the assessment of their biocompatibility as a preliminary step to their implantation *in vivo* [12-14]. Macrophages are essential cells of the innate immune response that can synthesize and release different cytokines upon stimulation by exogenous or endogenous factors. These cytokines can modulate macrophage functions and the expression of cell surface markers [43]. As mentioned above in the Introduction, pro-inflammatory M1 macrophages release characteristic cytokines and chemokines of this phenotype, including TNF-α, IL-6, IL-12, IL-1α, IL-1β, CXCL9, and CXCL10 [43,44]. IL-6 is produced by many cells, including osteoblasts, monocytes, macrophages and bone marrow mononuclear cells [45]. This cytokine and TNF-α play a crucial role in the inflammatory response, infection and stress [45-49]. To investigate the potential of MBG-75S particles to trigger an inflammatory response *in vivo*, their effect on the secretion of the pro-inflammatory cytokines TNF-α and IL-6 by RAW 264.7 and J774A.1 macrophages, two types of macrophage cell lines with different sensitivity, was analyzed after direct treatment with 1 mg/mL of MBG-75S particles.

The secretion of TNF-α was significantly increased by the contact of RAW 264.7 and J774A.1 macrophages with MBG-75S particles in comparison with the controls in the absence of material. However, a significant decrease of IL-6 secretion by these macrophages was detected after contact with MBG-75S particles (Table S3 in Supporting information). These two cytokines are coordinately produced by macrophages in response to different stimuli. However, it is important to note that TNF, unlike IL-6, has receptors on the macrophages themselves (TNF-R1 and TNFR-RII) [50] through which it produces activation of the transcriptional factor NFκB, thus mediating the cellular immune response [51, 52]. On the other hand, IL-6 also promotes the differentiation of cytotoxic T lymphocytes from both mature and immature T lymphocytes, therefore a decrease of this cytokine would be beneficial as it would be involved in lymphocyte activation, proliferation and differentiation towards a Th2 lymphocyte phenotype [53, 54].

### *3.4.* Effects of MBG-75S particles and NanoMBG-75S on proliferation, cell size, complexity, and intracellular content of reactive oxygen species (ROS) of RAW 264.7 and J774A.1 macrophages

After evaluating the adsorption of serum proteins on MBG-75S particles and their action on the production of proinflammatory cytokines by RAW 264.7 and J774A.1 macrophages, a complete study was carried out with these two cell lines, comparing the effects of MBG-75S particles (directly or deposited on transwell inserts) with those produced by NanoMBG-75S nanoparticles after their intracellular incorporation, in order to know if the distinct form of mesoporous bioactive glass administration would affect the response of macrophages in different ways. This work will contribute to the characterization of the innate immune response to this kind of biomaterials, prior to their application *in vivo*. Both, RAW 264.7 (ATCC® TIB-71™) and J774A.1 (ATCC® TIB-67™) cells are derived from a tumour in a male and female BALB/c mouse, respectively. J774A.1 cells are widely used in biomaterial biocompatibility studies, and they are semi-adherent cells. For this reason, it is to be expected that they are more sensitive to lack of anchorage in the presence of materials than RAW 264.7 cells [55].

Figures 4 and 5 show the effects of MBG-75S particles and NanoMBG-75S on proliferation, cell size, complexity, and intracellular content of reactive oxygen species (ROS) of RAW 264.7 and J774A.1 macrophages, respectively. For the analysis of all these cellular parameters, cells were cultured for 24 h in the presence of 1 mg/mL of MBG-75S, directly or indirectly by using transwell inserts (MBG T), and in the presence of 50 µg/mL of nanospheres (shown above in Scheme 1). Control cells without these materials were cultured in parallel.

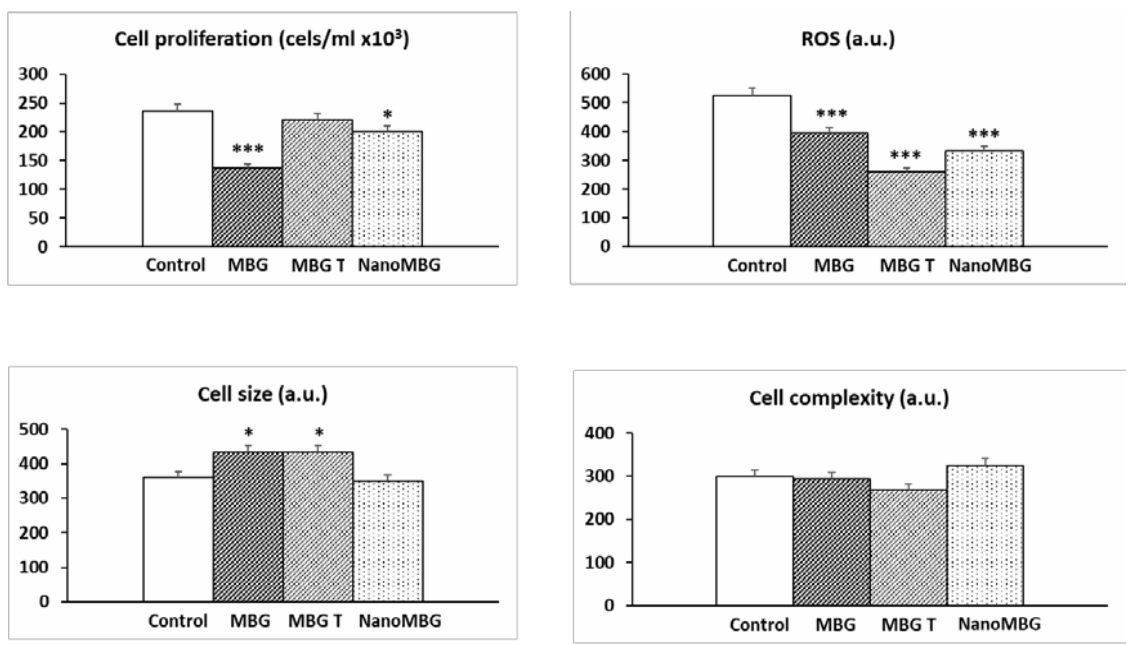

**Figure 3.** Effects of MBG-75S particles and NanoMBG-75S on proliferation, cell size, complexity, and intracellular content of reactive oxygen species (ROS) of RAW 264.7 macrophages. Statistical significance: * $p < 0.05$, *** $p < 0.005$ (comparison with control cells).

As it can be observed in Figure 3, the direct treatment of RAW 264.7 macrophages with MGB-75S particles induced a significant decrease on their proliferation. This effect may be due to cell damage caused by physical contact with this biomaterial that can induce the loss of anchorage, affecting both adhesion and division processes. However, in the case of indirect treatment with MBG-75S deposited on transwell inserts, no significant differences in proliferation were observed with respect to the control. It is important to note that the use of transwell inserts allows to evaluate only the action of the released ions by MBG-75S, avoiding the effects that could be generated by the physical presence of the material particles. On the other hand, the treatment with nanospheres slightly decreases RAW 264.7 proliferation. In previous studies we have observed that the incorporation of graphene oxide nanosheets alters the cell cycle phases of RAW 264.7 macrophages, decreasing the synthesis phase and thus affecting the proliferation of this type of macrophages [56].

Concerning the effects of these materials on cell size, only a slight but significant increase was observed after both direct and indirect treatments with MBG-75S particles. However, cell complexity was not affected by these MBG-75S treatments. This observed increase in cell size might be due to the silica network degradation as colloidal silicate fragments [57], which could also pass through the transwell membranes and be internalized by macrophages producing a cell size increase, as described for most mammalian adherent cells after nanoparticle uptake [35, 52].

When the intracellular content of reactive oxygen species (ROS) of RAW 264.7 macrophages was analyzed, it was observed that the different treatments with the biomaterials used in this study resulted in a decreased level of ROS. This fact indicates not only the absence of oxidative stress but also suggests a certain protective role of these materials against this type of stress, which, on the contrary, is induced by other biomaterials. Previous results of our group, obtained with MC3T3-E1 pre-osteoblasts, also demonstrated a decrease in intracellular ROS content after NanoMBG uptake [33,34].

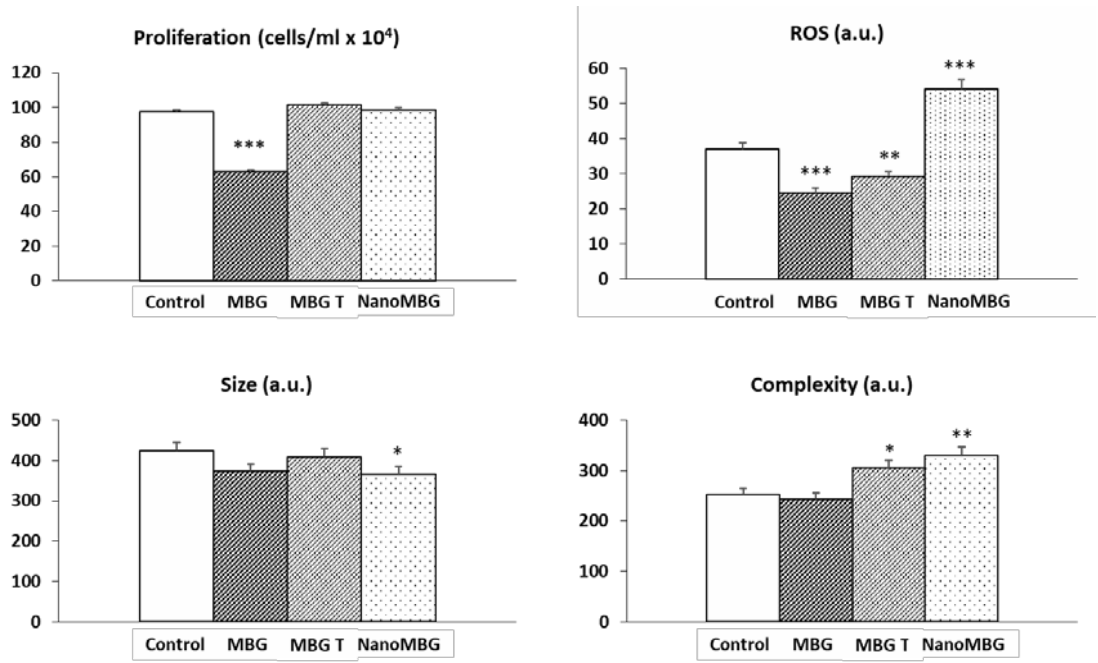

**Figure 4.** Effects of MBG-75S particles and NanoMBG-75S on proliferation, cell size, complexity, and intracellular content of reactive oxygen species (ROS) of J774A.1 macrophages. Statistical significance: * $p < 0.05$, *** $p < 0.005$ (comparison with control cells).

Figure 4 shows the effects of MBG-75S particles and NanoMBG-75S on proliferation, cell size, complexity, and intracellular content of reactive oxygen species (ROS) of J774A.1 macrophages. As it was observed with RAW 264.7 macrophages, after direct MBG-75S treatment, the proliferation of this type of macrophages significantly decreased with respect to the control ($p < 0.005$), probably due to the direct contact of the cells with the particles that can induce the loss of anchorage. However, no changes in proliferation were observed after treatment with MBG-75S deposited on transwell inserts or with NanoMBG-75S.

Regarding the effect of MBG-75S particles on the size and complexity of J774A.1 macrophages, only a slight but significant increase in complexity was observed after indirect treatment with this material (Figure 4), probably due to the release of ions through the transwell insert and/or to the uptake of colloidal silicate fragments, as indicated above [33,34,35]. The treatment of J774A.1 macrophages with NanoMBG-75S also produced an increase in cell complexity but was accompanied by a slight decrease in the size of these macrophages, suggesting a possible cellular retraction.

Cell size and complexity are properties that depend on different factors such as the state of the plasma membrane and several organelles, as well as the presence of granulated material within the cell [59, 60]. The measurement of these light scattering properties by flow cytometry is very useful to detect the changes that nanoparticles can induce in cells after their uptake, as indicated above [58,59,60].

On the other hand, although both direct and indirect treatments with MBG-75S particles produced a decrease of ROS in J774A.1 (Figure 4) and RAW 264.7 macrophages (Figure 3), a significant increase of intracellular content of these reactive species was observed after incubation of J774A.1 macrophages with 50 µg/mL of NanoMBG-75S. These results evidence the induction of oxidative stress by these nanospheres in this type of macrophages after their uptake. Considering that J774A.1 macrophages are more sensitive than RAW 264.7 cells, and with the aim of observing possible alterations of cell morphology caused by these biomaterials, confocal microscopy studies were carried out using phalloidin-rhodamine to stain the actin filaments of the cytoskeleton, DAPI to stain the nuclei and NanoMBG-75S labeled with FITC (FITC-NanoMBG-75S), as described in previous works [33,34]. For these studies, 1mg/mL of MBG-75S was deposited either in direct contact with cells or on transwell inserts, to evaluate only the action of the released ions. On the other hand, macrophages were also treated with 50 µg/mL of FITC-NanoMBG-75S to observe their intracellular incorporation. Control cultures without these biomaterials were performed in parallel.

As can be appreciated by the images in Figure 5, J774A.1 macrophages display the typical characteristics of this cell type after the indirect treatment with MBG-75S deposited on transwell inserts and after the incorporation of FITC-NanoMBG-75S. However, the direct contact with MBG-75S particles induced observable changes in cell shape. Thus, J774A.1 macrophages adopted a clearly more elongated morphology in agreement with the M2 reparative phenotype of macrophages (Figure 3, MBG-75S) as it has been described by other authors as a characteristic of reparative macrophages [43].

In a previous work, we analyzed the effects of different doses of MBG-75S particles on the morphology of human Saos-2 osteoblasts, observing the typical osteoblast characteristics in the presence of 0.5 and 1 mg/mL of this material. However, higher MBG-75S doses induced cell morphology alterations and revealed the existence of apoptosis [25].

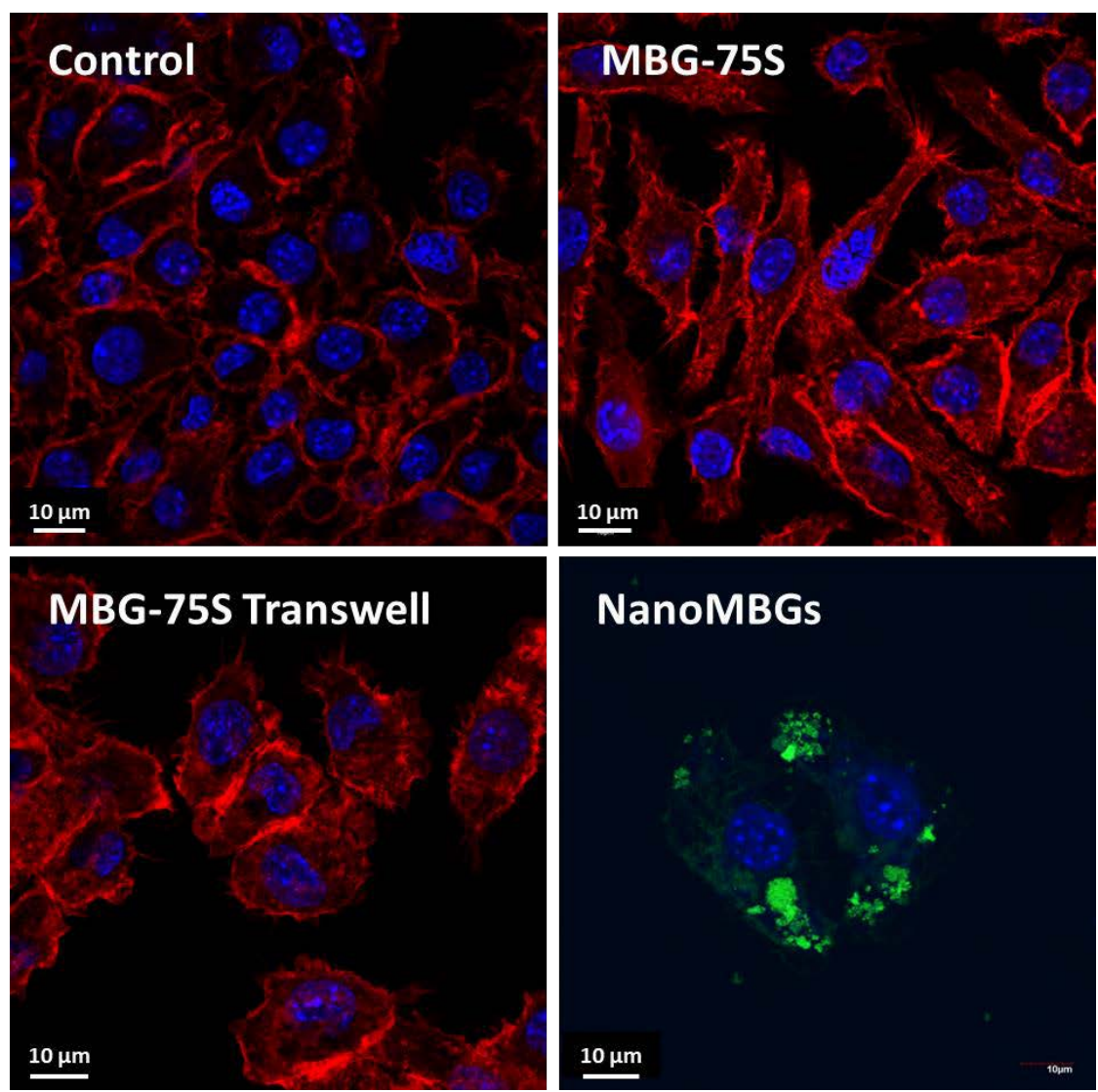

**Figure 5.** Confocal microscopy images of J774A.1 macrophages after 24 hours of treatment with 1mg/mL of MBG-75S, either in direct contact with cells (MBG-75S) or deposited on transwell inserts (MBG-75S Transwell), and after incorporation of FITC-NanoMBG-75S (50 µg/mL). Nuclei were stained with DAPI (blue), F-actin filaments were stained with rhodamine-phalloidin (red) and FITC-NanoMBG-75S are observed in green.

*3.5.* **Effects of MBG-75S particles and NanoMBG-75S on polarization of RAW 264.7 and J774A.1 macrophages towards pro-inflammatory M1 and reparative M2 phenotypes.**

Recent *in vitro* studies with RAW 264.7 macrophages treated with MBG-75S particles [25] and with NanoMBG-75S [34] showed that these biomaterials do not induce macrophage polarization towards the proinflammatory M1 phenotype. The present work has allowed us to further study the response of macrophages to these mesoporous materials and to compare for the first time their effects in the form of particles and nanospheres, using two types of macrophages with different sensitivity.

The action of MBG-75S particles and NanoMBG-75S on the polarization of RAW 264.7 macrophages towards pro-inflammatory (M1) and reparative (M2) phenotypes was evaluated by flow cytometry through the expression of CD80 and CD206, as M1 and M2 markers, respectively.

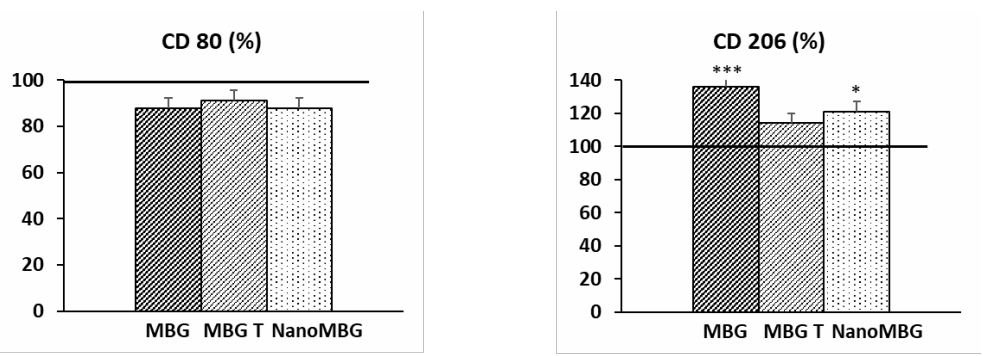

**Figure 6.** Effects of MBG-75S particles and NanoMBG-75S on polarization of RAW 264.7 macrophages. Percentage of proinflammatory M1 macrophages (CD80$^+$ cells) and reparative M2 macrophages (CD206$^+$ cells) after 24 h of treatment with 1 mg/mL of MBG-75S deposited either in direct contact with cells (MBG) or on transwell inserts (MBG T) and after incorporation of NanoMBG-75S (50 µg/mL). All the values are shown as percentages referred to the corresponding control value of each parameter taken as 100% (horizontal line). Statistical significance: * $p < 0.05$, *** $p < 0.005$ (comparison with control cells).

As shown in Figure 6, the treatment of RAW 264.7 macrophages with MBG-75S particles (directly or deposited on transwell inserts) and NanoMBG-75S, clearly decreased the percentage of M1 macrophages (CD80$^+$ cells) after 24h of culture, although this decrease was not statistically significant. On the other hand, these treatments clearly induced the macrophage polarization towards the M2 repair phenotype by increasing the expression of CD206, a representative marker of this phenotype as indicated above. The positive effects on CD206 expression after direct contact with MBG-75S particles and in the presence of nanospheres were significant and more pronounced than those obtained after indirect treatment with the particles deposited on transwell inserts. Previous studies with RAW 264.7 macrophages treated with NanoMBG-75S also evidenced a decrease of the M1 phenotype [34]. On the other hand, we have observed a similar macrophage behavior after treatment with graphene oxide nanoparticles [48].

In order to better understand the effects of these mesoporous materials on the polarization of macrophages towards M1 or M2 phenotypes, we carried out similar studies with J774A.1 macrophages but including the expression of CD163 as an extra M2 phenotype marker [46,49].

As it can be observed in Figure 7, in this cell line, the direct treatment with MBG-75S particles and NanoMBG-75S also significantly decreased the percentage of M1 pro-inflammatory macrophages (CD80$^+$ cells) after 24h. In contrast, MBG-75S deposited on transwell inserts (MBG T) did not modify the percentage of these cells after 24 h of indirect treatment.

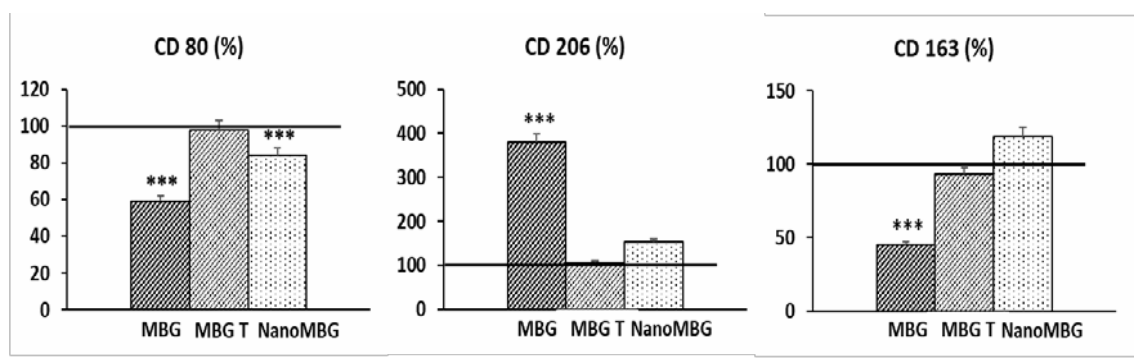

**Figure 7.** Effects of MBG-75S particles and NanoMBG-75S on polarization of J774A.1 macrophages. Percentage of proinflammatory M1 macrophages (CD80$^+$ cells) and reparative M2 macrophages (CD206$^+$ and CD163$^+$ cells) after 24 h of treatment with 1 mg/mL of MBG-75S deposited either in direct contact with cells (MBG) or on transwell inserts (MBG T) and after incorporation of NanoMBG-75S (50 µg/mL). All the values are shown as percentages referred to the corresponding control value of each parameter taken as 100% (horizontal line). Statistical significance: * $p < 0.05$, *** $p < 0.005$ (comparison with control cells).

On the other hand, the direct treatment of J774A.1 macrophages with MBG-75S particles induced a significant and very pronounced increase of CD206 expression. However, the treatment of these macrophages with MBG-75S particles deposited on transwell inserts or with NanoMBG-75S did not induce significant changes of this M2 marker.

Considering that J774A.1 macrophages are more sensitive than RAW 264.7 cells and with the aim of observing possible changes in other markers of the M2 reparative phenotype, the expression of CD163 was evaluated as indicated above, after the different treatments with these biomaterials. As show in Figure 7, the treatment with

MBG-75S particles in direct contact with J774A.1 cells induced a significant decrease of CD163+ cells. However, these particles deposited on transwell inserts and NanoMBG-75S did not induce significant changes in the expression of this M2 marker by J774A.1 macrophages. CD163 is a scavenger receptor highly expressed on resident tissue macrophages, acting as a receptor for hemoglobin-haptoglobin complexes and mediating interactions between macrophages and erythroblasts. It has also been demonstrated that CD163 can function as a macrophage receptor for bacteria acting as an innate immune sensor and inducing local inflammation [61]. Considering this possible action of CD163 in tissue resident macrophages, its decrease could be a beneficial effect produced by MBG-75S particles in the case of implantation.

The results obtained in this study with RAW 264.7 and J774A.1 macrophages demonstrate that MBG-75S particles and NanoMBG-75S decrease the percentage of M1 cells, promoting the shift of the M1/M2 balance towards M2 reparative phenotype, evidencing an appropriate immune response to this kind of biomaterials.

Moreover, the very pronounced increase of CD206 expression induced by MBG-75S particles on J774A.1 macrophages can be associated with changes in cell shape related to the M2 reparative phenotype [43], that were also observed in the present study when these macrophages were directly treated with MBG-75S particles.

The results obtained in this work are also in agreement with previous work, in which it was observed that NanoMBG-75S did not affect the balance of spleen cell subsets, or the production of intracellular or secreted pro- and anti-inflammatory cytokines (TNF-α, IFN-γ, IL-2, IL-6, IL-10) by activated T, B and dendritic cells (DC) [58].

**Conclusions**

This comparative study demonstrates that MBG-75S particles and NanoMBG-75S nanospheres show an excellent behavior respect to innate immune response, as they allow the appropriated development and function of RAW 264.7 and J774A.1 macrophages, two cell lines with different sensitivity, and do not induce polarization towards the M1 pro-inflammatory phenotype in these cells. In contrast, particles and nanoparticles of mesoporous bioactive glass induce a very pronounced increase of CD206 expression, a specific marker of the M2 reparative phenotype, on RAW 264.7 and J774A.1 macrophages. Therefore, considering that both mesoporous biomaterials offer the possibility of loading drugs into their pores, the results obtained indicate their high potential for use as drug-delivery systems in bone repair and osteoporosis treatments without triggering an adverse inflammatory response.

**ACKNOWLEDGEMENTS**

This study was supported by research grants from the Ministerio de Economía y Competitividad, Agencia Estatal de Investigación (AEI) and Fondo Europeo de Desarrollo Regional (FEDER) (MAT2016-75611-R AEI/FEDER, UE). MVR acknowledges funding from the European Research Council (Advanced Grant VERDI; ERC-2015-AdG Proposal No. 694160). L.C. is grateful to the Universidad Complutense de Madrid for a UCM fellowship.The authors wish to thank the staff of the ICTS Centro Nacional de Microscopia Electrónica (Spain) and the Centro de Citometría y Microscopía de Fluorescencia of the Universidad Complutense de Madrid (Spain) for the assistance in the electron microscopy, flow cytometry and confocal microscopy studies.

**Author Contributions:** Conceptualization, M.J.F., D.A. and M.T.P.; methodology, M.J.F., L.C., M.O., M.G.-D., P.A., A.P.-M., D.A. and M.T.P.; validation, M.J.F., D.A. and M.T.P.; formal analysis, L.C., M.G.-D., P.A.; investigation, M.J.F., M.O., D.A. and M.T.P.; methodology, M.J.F., L.C., M.O., M.G.-D., P.A., A.P.-M., D.A. and M.T.P; resources, M.V., M.T.P. and D.A.; data curation, M.J.F., D.A. and M.T.P.; writing—original draft preparation, M.J.F., D.A. and M.T.P.; writing—review and editing, M.J.F., D.A. and M.T.P.; visualization, M.J.F., D.A. and M.T.P.; supervision, M.J.F., D.A. and M.T.P.; project administration, M.V., D.A. and M.T.P.; funding acquisition, M.V., D.A. and M.T.P. All authors read and agreed to the published version of the manuscript.

**Declaration of interests**

☒ The authors declare that they have no known competing financial interests or personal relationships that could have appeared to influence the work reported in this paper.

☐The authors declare the following financial interests/personal relationships which may be considered as potential competing interests: